\documentclass{article}
\usepackage{graphicx,color,amsmath, amssymb,multirow}
\usepackage[english]{babel}
\usepackage{t1enc}
\usepackage[latin2]{inputenc}
\begin{document}
\title{Dilepton creation based on an analytic hydrodynamic solution}

\author{M.~Csan\'ad\footnote{The author acknowledges the support of the OTKA grant 101438}, L.~Krizs\'an\\
\small{E\"otv\"os University, Department of Atomic Physics, H-1117 Budapest, Hungary}}

\maketitle

\begin{abstract}
High-energy collisions of various nuclei, so called ``Little Bangs'' are observed at various experiments of
heavy ion colliders.  The time evolution of the strongly interacting quark-gluon plasma created in heavy
 ion collisions can be described by hydrodynamical models. After expansion
and cooling, the hadrons are created in a freeze-out. Their distribution describes the final state of
this medium. To investigate the time evolution one needs to analyze penetrating probes, such as direct
photon or dilepton observables, as these particles are created throughout the evolution of the medium.
In this paper we analyze an 1+3 dimensional analytic solution of relativistic hydrodynamics, and
we calculate dilepton transverse momentum and invariant mass distributions. We investigate the dependence of
dilepton production on time evolution parameters, such as emission duration and equation of state. Using parameters
from earlier fits of this model to photon and hadron spectra, we compare our calculations to measurements as well.
The most important feature of this work is that dilepton observables are calculated from an exact, analytic, 1+3D solution
of relativistic hydrodynamics that is also compatible with hadronic and direct photon observables.
\end{abstract}

\section{Introduction}
The interest in relativistic perfect fluid hydrodynamics grew in the last decade due to the discovery 
of the strongly interacting quark gluon plasma: it is well known that the medium created in
high energy heavy ion collisions is an almost perfect fluid, in particular the collective phenomena in
soft hadron production can be most successfully described by hydrodynamic
models~\cite{Adcox:2004mh}. Hydrodynamic models aim to describe the dynamics of the medium
created in heavy ion collisions, and investigate the relation between the initial state, the final state
and the equation of state. Many solve the equations of hydrodynamics numerically, but there
exist a few exact, analytic solutions.  Goal of this paper is to obtain analytic results on dilepton distributions
based on an exact hydrodynamic solution. In case of numerical simulations, one may take more realistic
effects into account, but one loses the analytic understanding of the connection of input parameters and final results.
This understanding can however better be maintained when using exact, analytic solutions.

There was a long search for such exact, analytic solutions of relativistic hydrodynamics, and
only few applicable ones were found. Most of these are 1+1 dimensional, and few truly 1+3 dimensional
solutions exist. In this paper we extract observables from the relativistic, ellipsoidally symmetric solution of
Ref.~\cite{Csorgo:2003ry}. Hadronic observables from this solution were calculated in Ref.~\cite{Csanad:2009wc}
and compared to RHIC data successfully. Obtained model parameters describe the final state (the hadronic freeze-out),
where hadrons are created. The time evolution of the medium can be probed by penetrating probes: leptons and
photons, created throughout the evolution. Direct photon observables, based on the before mentioned model,
were calculated in Ref.~\cite{Csanad:2011jq}, and they were also successfully compared to RHIC data.
In this paper we calculate transverse momentum distribution and invariant mass distribution of dileptons
from the solution of Ref.~\cite{Csorgo:2003ry}: this is the ``missing link'', the most important quantity
not yet calculated via a model based on an exact, analytic 1+3D relativistic hydro solution.

\section{Perfect fluid hydrodynamics}
The equations of hydrodynamics describe the local conservation of a given charge ($n$) and local
conservation of energy-momentum
density ($T^{\mu\nu}$). The fluid is perfect if the energy-momentum tensor is diagonal in the local rest frame,
i.e.\  viscosity and heat conduction are negligible. This can be assured if $T^{\mu\nu}$ is chosen as
\begin{align}
T^{\mu\nu}=(\epsilon+p)u^\mu u^\nu-pg^{\mu\nu},
\end{align}
where $\epsilon$ is energy density, $p$ is pressure and $g^{\mu\nu}$ is the metric tensor, diag$(-1,1,1,1)$, while
 $x^{\mu} = (t, r_x, r_y, r_z)$ is a given point in space-time, $\tau = \sqrt{t^2-r^2}$ the coordinate proper-time,
 $\partial_{\mu} = \frac{\partial}{\partial x^{\mu}}$ is the derivative versus space time, while
 $p^{\mu} = (E, p_x, p_y, p_z)$ is the four-momentum.

The conservation equations are closed by the equation of state (EoS), which defines
relationship between energy density $\epsilon$ and pressure $p$. Usually $\epsilon = \kappa p$ is chosen,
where $\kappa$ may depend on temperature $T$, and the first relativistic exact solutions with temperature
dependent $\kappa$ were recently found in Ref~\cite{Csanad:2012hr}. In this paper we however use a solution
with constant $\kappa$. It is important to see that in this case $\kappa = 1/c_s^2$, with $c_s$ being the speed 
of sound. Temperature can then be defined based on entropy density, energy density and pressure.

If a solution of the above partial differential equations is given, the phase-space distribution 
can be expressed by a (Boltzmann-)Jüttner-distribution of
\begin{align}
f(k,x)\propto \exp\left[-\frac{k_\mu u^\mu}{T(x)}\right].\label{e:juttner}
\end{align}
The hadronic observables can be extracted from a solution via this phase-space distribution at the
freeze-out, based on the choice of an emission surface and Cooper-Frye like emission function. This
distribution corresponds to the hadronic final state or source distribution (see details about such a
calculation in Ref.~\cite{Csanad:2009wc}). In case of dileptons or photons, one has to follow a different path.

An important result for hydrodynamic models is, that because hadrons are created at the quark-hadron transition,
hadronic observables do not depend on the initial state or the dynamical equations (equation of state) separately,
just through the final state~\cite{Csanad:2009sk}. Thus same hadronic final state can be achieved with different
equations of state or initial conditions. We may fix the final state from hadronic data, but we need
penetrating probes, such as photon or lepton data to investigate the equation of state or the initial state.
Photon creation is sensitive to the whole time evolution, thus both to initial conditions and equation of state as well, 
as discussed in Ref.~\cite{Csanad:2011jq}, where direct photon observables were calculated and compared
to data. Dilepton creation is also sensitive to the whole time evolution, as we will see in the next sections.

\section{Dilepton creation from hydro evolution}
Many calculations exist on the nature of dilepton production in a hadron gas or in QGP, based the thermal
expectation value of the electromagnetic current-current correlators, see e.g. those in Refs.~\cite{Dusling:2006yv,Ghosh:2010wt}.
Dilepton emission from based on a hydrodynamic solution can be calculated similarly to
Refs.~\cite{Kajantie:1986cu,Asakawa:1993kb,vanHees:2006ng,Renk:2006qr,Ruppert:2007cr,Alam:2009da,Song:2010fk}. Hydrodynamics
in these works is utilized in form of a full-blown numerical simulation or approximative solutions. Our goal is, however,
to obtain an analytic description of the dilepton distributions based on an analytic solution of hydrodynamics.
In order to do this, we start from the dilepton source given as
\begin{align}
\frac{dN}{d^4x} = \int d^3 k_1 d^3 k_2 f(k_1,x) f(k_2,x) v_\textnormal{rel} \sigma
\end{align}
with $k_1$ and $k_2$ being the momenta of the two particles creating the 
dilepton pair, $f(k_i,x)$ the Jüttner-distribution, $v_\textnormal{rel}$ is the relative
velocity of the incoming pair, and $\sigma$ is the 
production cross-section of the given process. This is based on the well-known
relation from nuclear physics, where the rate of a given process is proportional
the density of the involved particles, the cross-section of the process and the
average velocity: rate($A+B\rightarrow X$)$=n_An_B\langle \sigma_{A+B\rightarrow X} v\rangle$.
The assumption in this calculation is, that dileptons are created in annihilation--decay-like
processes: $q\bar{q}$ is the incoming pair the quark-gluon phase, and $\pi^+ \pi^-$ in
the hadron gas phase. Then the dilepton creation happens through a quasireal photon or
a vector meson.

The actual calculation of the creation rate goes as follows. First, we express the relative velocity 
in the pair-comoving frame in the following invariant form:
\begin{align}
v_\textnormal{rel} = \frac{M^2}{2E_1E_2}\sqrt{1-\frac{4m^2}{M^2}}
\end{align}
with $M^2$ being the dilepton invariant mass squared and $m$ the mass
of the two incoming particles (pions or quarks) and $E_{1,2}$ their energy. Then we switch from 
the $k_1$ and $k_2$ to $P=k_1+k_2$ and $k=(k_1-k_2)/2$, and obtain in the pair center of mass system
\begin{align}
\frac{d^3 k_1}{E_1} \frac{d^3 k_2}{E_2} = \frac{d^3P}{E} \frac{4d^3k}{M}
= \frac{d^3P}{E} \frac{1}{16}\sqrt{1-\frac{4m^2}{M^2}} MdM d\Omega,
\end{align}
with $E=E_1+E_2$ and $\Omega$ the solid angle corresponding to $P$; finally, the dilepton source is 
\begin{align}\label{e:dilepton_source}
\frac{dN}{d^4x} = \int \frac{d^3 P}{E} \frac{1}{16}\sqrt{1-\frac{4m^2}{M^2}} 
MdM d\Omega f(k_1,x) f(k_2,x) \frac{M^2}{2}\sqrt{1-\frac{4m^2}{M^2}} \sigma.
\end{align}

Next, we have to assume a cross-section for dilepton creation.
The color-averaged $q\bar q \rightarrow l^+ l^-$ cross-section for a given
flavor can be calculated as~\cite{Kajantie:1986dh}
\begin{align}
\sigma = \frac{4\pi\alpha^2}{9M^2} e_q^2\frac{1+2m_q^2/M^2}{\sqrt{1-4m_q^2/M^2}}
\end{align}
with $M^2$ being the dilepton invariant mass squared, $e_q$ the charge of 
the given quark (in units of $e$), and $m_q$ its mass. This then has to be summed
up for the used flavors ($u,d,s$ in this case).

The same cross-section for dilepton production in a pion gas can be calculated as
\begin{align}
\sigma = \frac{4\pi\alpha^2}{3M^2} |F(M^2)|^2 \sqrt{1-4m_\pi^2/M^2}
\end{align}
where again $M^2$ is the dilepton invariant mass squared, $m_\pi$ the pion mass,
and $F(M^2)$ is the electromagnetic form factor of pions, summed up for
a range of exchangeable particles ($\rho$ and its excitations $\rho'$ and $\rho''$,
as well as $\omega$ or $\phi$):
\begin{align}
|F(M^2)|^2 = \sum \frac{N_i m_i^4}{(m_i^2-M^2)^2 + m_i^2 \Gamma_i^2}
\end{align}
where $N_i$ is a relative normalization factor (for which, in case of $\rho$, $\rho'$
and $\rho''$ the values of 1, $8.02\times 10^{-3}$ and $5.93\times 10^{-3}$ are
taken usually~\cite{Asakawa:1993kb}, and these factors have to be tuned to
experimental data for other particles as well), $m_i$ is the mass of the exchanged
particle and $\Gamma_i^2$ its width. One may use the vacuum mass and width
of the given particles, or their in-medium value, if there is a hint at an in-medium
spectral function modification. It is the case for high energy heavy ion collisions,
however, we will stick to the vacuum values, taken from Ref.~\cite{Beringer:1900zz}.

In order to obtain the mass- and average momentum dependence of dilepton
creation, we assume a Jüttner-like distribution of the incoming particles, as
given also in Eq.~(\ref{e:juttner}):
\begin{align}
f(k,x)d^3k = \frac{gd^3k}{(2\pi)^3}e^{-k^\mu u_\mu(x)/T(x)}.
\end{align}
This means that we assume a thermalized medium containing quarks, or
a (still) thermalized pion gas. Dileptons can be created in both media, the
main difference will be the cross-section (the production mechanism described in the above
sections), the temperature of the medium of the `incoming' particles, and the duration
of the emission (as well as the expanding size of the particle emitting region).
In Eq.~(\ref{e:dilepton_source}) $f(k_1,x)f(k_2,x)$ shows up, but this clearly
equals $f(P,x)$, if $P=k_1+k_2$ is the total momentum (note, that the original formula
of  Eq.~(\ref{e:dilepton_source}) neglects quantum-statistical and other correlations
of the pairs, which is a good approximation, as this effect is smeared out by integrating
on the relative momentum). Taking this into account,
we get the following general result
\begin{align}
\frac{dN}{dyMdMd^2P_t} = \frac{g^2\pi}{16(2\pi)^5}M^2\left(1-\frac{4m^2}{M^2}\right) 
\sigma \int e^{-P^\mu u_\mu(x)/T(x)} d^4x.\label{e:dilepton_emission}
\end{align}
(where $m$ is the mass of the incoming particle, $m_\pi$ or $m_q$, so in fact the above is a separate
formula for the QGP and the pion gas case). Next, we utilize this and a given solution
of hydrodynamics to calculate dilepton distributions.

\section{Dilepton creation in the analyzed solution}
The analyzed 1+3D relativistic solution~\cite{Csorgo:2003ry} assumes self-similarity and ellipsoidal symmetry.
Thermodynamical quantities (in particular the temperature, as discussed below) are constant on the surface of
expanding ellipsoids, given by constant values of
\begin{align}
s=\frac{r_x^2}{X(t)^2}+\frac{r_y^2}{Y(t)^2}+\frac{r_z^2}{Z(t)^2},\label{e:scalevar}
\end{align}
where $X(t)$, $Y(t)$, and $Z(t)$ are time dependent scale parameters (axes of the $s=1$ ellipsoid),
only depending on the time. Note that in this paper, we use $X(t)=Y(t)$ in order to obtain a simple
result. This is justified as azimuthal asymmetry is integrated out in the investigated observables.
The velocity-field describes a spherically symmetric Hubble-expansion:
\begin{align}
u^\mu (x) = \frac{x^\mu}{\tau},
\end{align}
and $\dot X(t), \dot Y(t), \dot Z(t)$ have to be constant to satisfy the equations of hydrodynamics.

The temperature distribution in this model $T(x)$ is
\begin{align}
T(x)=T_0\left(\frac{\tau_0}{\tau}\right)^{3/\kappa} e^{bs/2},\label{e:temp}
\end{align}
where $\tau_0$ is an arbitrary time when the temperature is $T_0$. This is usually chosen to be
 the time of the freeze-out and thus $T_0$ is the central freeze-out temperature (i.e.
$T_0 = \left.T\right|_{s=0,\tau=\tau_0}$), while $s$ is the above scaling variable and $b$ is a
parameter defining the temperature gradient. If the fireball is the hottest in the center, then $b<0$.
The above outlined solution may be applied to the time evolution of the strongly interacting QGP,
but also to a thermalized hadron gas.

If we plug this solution in Eq.~(\ref{e:dilepton_emission}), we have to integrate the phase space 
distribution with respect to $x$. To perform this integration we will use a second order saddlepoint
approximation.\footnote{In case of
a saddlepoint-type of integration is, we have an $\int f g$ type of expression, with $f$ being a narrow
distribution around $x_0$ and $g$ a slowly changing function. The second order saddle-point approximation
can then be given as $g(x_0) f(x_0) \sqrt{\pi}\Delta f$ with $\Delta f$ being the width
of the narrow distribution $f$.}
In this approximation the point of maximal emissivity (the maximum of the $f(k,x)$ Jüttner
distribution in $x$) is
\begin{align}\label{e:r0}
r_{0,x} = \rho_x t \frac{P_x}{E}\\
r_{0,y} = \rho_y t \frac{P_y}{E}\\
r_{0,z} = \rho_z t \frac{P_z}{E}
\end{align}
while the widths of the particle emitting source distribution are
\begin{align}\label{e:R}
R_x^2 = \rho_x \left( \frac{t}{\tau_0}\right)^{-3/\kappa+2} \tau_0^2 \frac{T_0}{E}\\
R_y^2 = \rho_y \left( \frac{t}{\tau_0}\right)^{-3/\kappa+2} \tau_0^2 \frac{T_0}{E}\\
R_z^2 = \rho_z \left( \frac{t}{\tau_0}\right)^{-3/\kappa+2} \tau_0^2 \frac{T_0}{E}
\end{align}
where we introduced the auxiliary quantities
\begin{align}\label{e:rho} 
\rho_x = \frac{\kappa}{\kappa -3-\kappa\frac{b}{\dot{X_0^2}}}\\
\rho_y = \frac{\kappa}{\kappa -3-\kappa\frac{b}{\dot{Y_0^2}}}\\
\rho_z = \frac{\kappa}{\kappa -3-\kappa\frac{b}{\dot{Z_0^2}}}
\end{align}
where again $\kappa=c_s^{-2}$ is describing the EoS. The source widths depend clearly on time, as the system is expanding.
Note that we assume $\dot{X_0}=\dot{Y_0}$ in this paper, as noted after Eq.~(\ref{e:scalevar}). The source then looks like
\begin{align}
e^{-P^\mu u_\mu(x)/T(x)} = e^{C-\frac{(r_x-r_{x,0})^2}{2R_x^2}
-\frac{(r_y-r_{y,0})^2}{2R_y^2}-\frac{(r_y-r_{y,0})^2}{2R_y^2}}\textnormal{, with}\\
C=-\frac{E}{T_0}\left(\frac{t}{\tau_0}\right)^{3/\kappa}
\left(E^2-(\rho_x P_x +\rho_y P_y + \rho_z P_z)/2\right).
\end{align}
After integrating on the spatial coordinates, we get
\begin{align}
\int e^{-P^\mu u_\mu(x)/T(x)} d^3x = e^C (2\pi)^{3/2} \sqrt{\rho_x\rho_y\rho_z}
\left(\frac{T_0\tau_0^2}{E}\right)^{3/2}\left(\frac{t}{\tau_0}\right)^{-2/\kappa+2}.
\end{align}
At midrapidity and after integration on the azimuthal angle of the momenta our result is
\begin{align}
&\exp\left[-\frac{1}{T_0\sqrt{M^2+P_t^2}}\left(\frac{t}{\tau_0}\right)^\frac{3}{\kappa}
\left(M^2-P_t^2-\frac{\rho_x+\rho_y}{4}P_t^2\right)\right](2\pi)^\frac{5}{2}\sqrt{\rho_x\rho_y\rho_z}\;\times\nonumber\\
&\left(\frac{T_0\tau_0^2}{\sqrt{M^2+P_t^2}}\right)\left(\frac{t}{\tau_0}\right)^{-\frac{3}{\kappa}+3}
I_0\left(\frac{P_t^2}{T_0\sqrt{M^2+P_t^2}}\left(\frac{t}{\tau_0}\right)^{3/\kappa}P_t^2
\frac{\rho_y-\rho_x}{4}\right).
\end{align}
Now we have to integrate on time, and assuming $\rho_x=\rho_y$ (i.e. neglecting azimuthal asymmetry\footnote{Note,
that the experimental dilepton observables are azimuthally integrated, so azimuthal asymmetry does not have to be
taken into account.}), our result on the dilepton source is
\begin{align}\label{e:dilepton_result}
\frac{dN}{MdMP_tdP_t} &= \frac{g^2}{16(2\pi)^{5/2}}M^2\left(1-\frac{4m^2}{M^2}\right) 
\sigma \sqrt{\rho_x\rho_y\rho_z}\tau_0^4\;\times\nonumber\\
& \left(\frac{T_0\tau_0^2}{\sqrt{M^2+P_t^2}}\right)^{3/2}\kappa A^{\frac{3}{2}-\frac{4\kappa}{3}}
 \left.\Gamma\left(\frac{4\kappa}{3}-\frac{3}{2}; A \xi^\frac{3}{\kappa}\right)\right|_{\xi=t_i/t_0}^{\xi=t_f/t_0}
\end{align}
where we introduced the following constant:
\begin{align}
A = \frac{M^2-P_t^2\left(1+\frac{\rho_x+\rho_y}{4}\right)}{T_0\sqrt{M^2+P_t^2}},
\end{align}
and $\xi=t/t_0$ is the time-variable divided by the $t_0$, the ``fixed point'' of the time integration. We will
choose this to correspond to the time of the quark-hadron transition (sometimes called the hadronic freeze-out),
thus $t_0=t_{\rm fo}$, and $T_0=T_{\rm fo}$ is the temperature of this transition. The time limits of the
integration for QGP are then $[t_{\rm ini},t_{\rm fo}]$, while
for the hadron gas phase they are $[t_{\rm fo},t_{\rm final}]$. Here $t_{\rm ini}$
is the initial time of dilepton production, and $T_{\rm ini}$ is the associated temperature, and similarly 
 $t_{\rm final}$ is the time when dilepton production in the hadron gas stops (one may call it
 a kind of kinetic freeze-out), and $T_{\rm final}$ is  the corresponding temperature.
 Based on Eq.~(\ref{e:temp}), one can also express $\xi$
as a function of the central temperature (since $t=\tau$ here): $\xi=(T_{\rm fo}/T(s=0,t))^{\kappa/3}$. We will
use the temperature range to determine the integration limits when comparing the experimental data.

Given Eq.~(\ref{e:dilepton_result}), one substitutes the given cross-section $\sigma$
in the above formula, and obtains the dilepton distribution based on the given production mechanism.
Eq.~(\ref{e:dilepton_result}) represents the main result of this paper: an analytic formula for the dilepton
invariant mass and transverse momentum distribution\footnote{The distribution given in Eq.~\ref{e:dilepton_result}
depends on both $M$ and $P_t$, and it has to be integrated out numerically in order to obtain for example
$dN/dM$}, for various production mechanisms. In the next section
we analyze this result quantitatively, mainly its dependence the model parameters.

\section{Quantitative analysis of our results}
Let us calculate the dilepton production from a thermal quark-gluon plasma type of medium, discovered
at RHIC, and from a hadron gas, dominating dilepton production at SPS. Model parameters in such a 
calculation can be based on comparison of the model to hadron~\cite{Csanad:2009wc} and photon
production~\cite{Csanad:2011jq}. In these fits to $\sqrt{s_{NN}}=200$ GeV RHIC Au+Au data, we
got $T_{\rm fo}=204$ MeV for the central (maximal)
temperature at the freeze-out, -0.34 or -0.84 (depending on centrality) for the transverse expansion over temperature
gradient $\dot{X_0^2}/b$ and $\dot{Y_0^2}/b$ (similarly to Eq.~(\ref{e:rho}), results don't depend on
expansion or temperature gradient separately, just through their ratio), $\tau_0=7.7$ fm$/c$ for the
freeze-out proper time. In this section we analyze the quantitative dependence of dilepton creation (invariant
mass and transverse momentum distributions) on the most important parameters of $\kappa$ (the
Equation of State) and the dilepton creation time interval, using values for the other parameters mentioned
above.  In the next section we will then compare our calculations to RHIC and SPS data.

\begin{figure}
 \begin{center}
 \includegraphics[height=0.49\textwidth, angle=270]{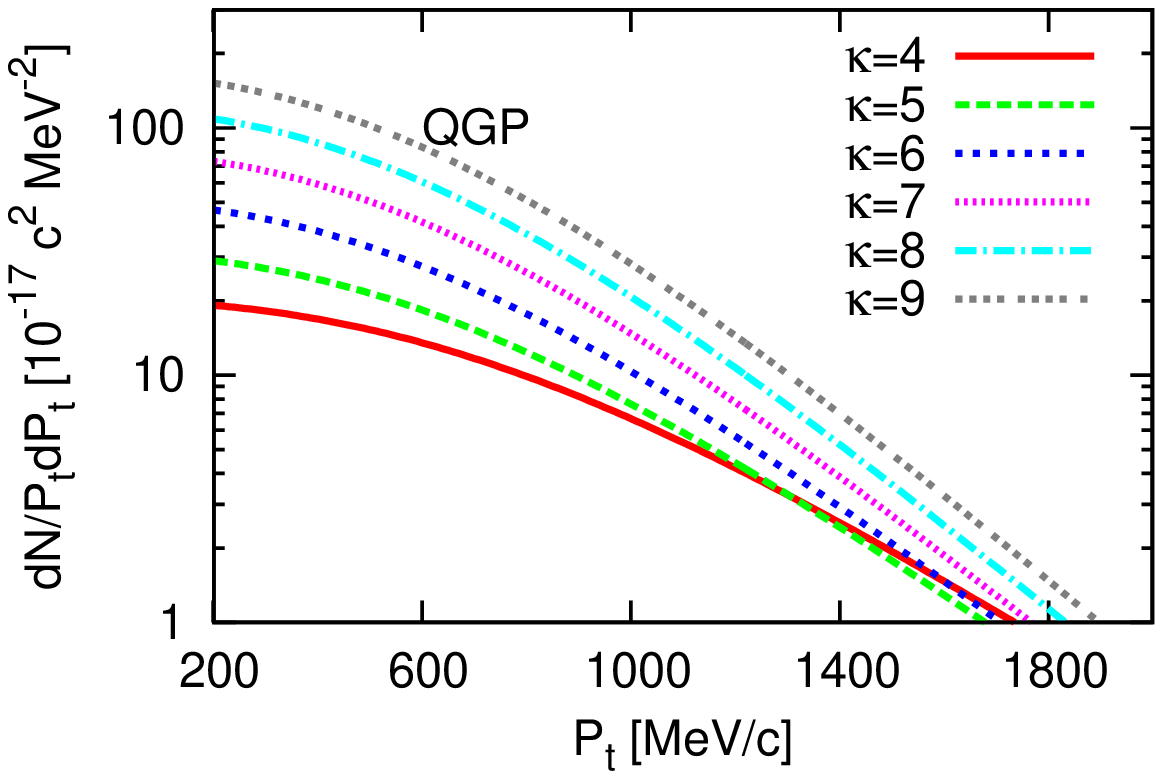}
 \includegraphics[height=0.49\textwidth, angle=270]{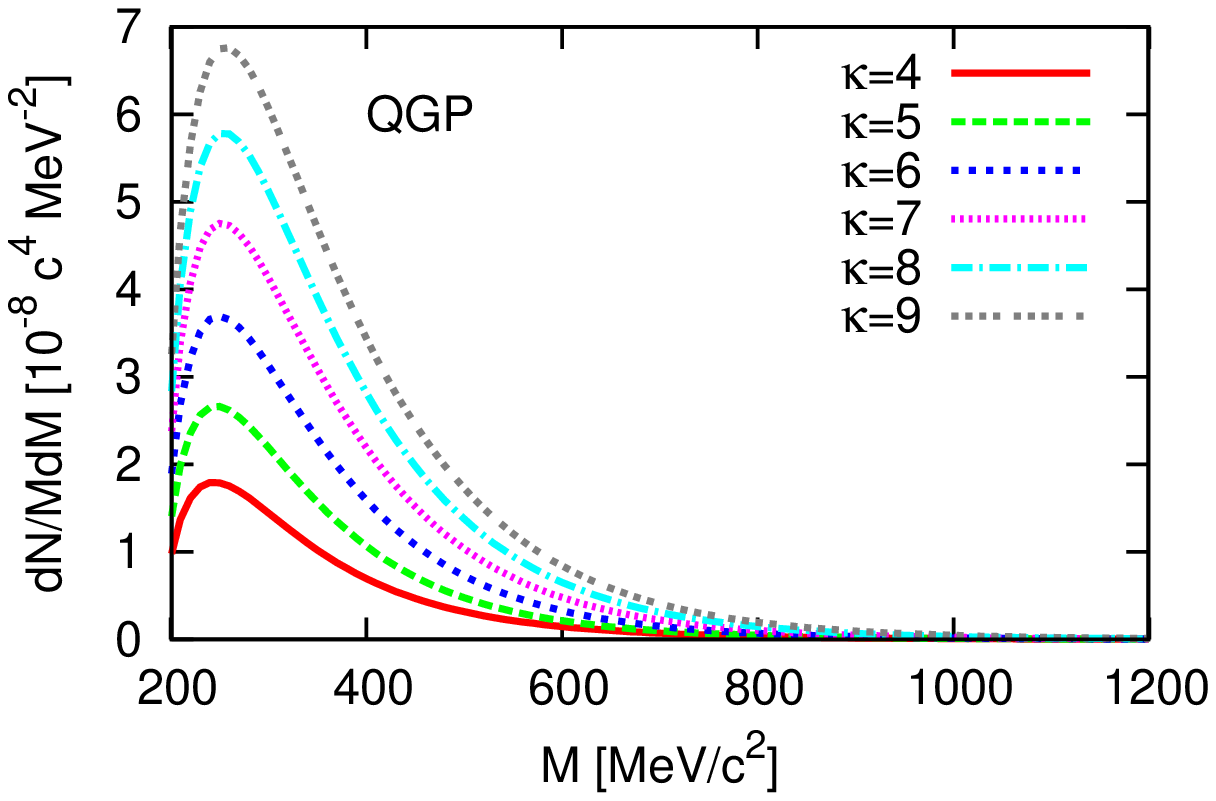}\\
 \includegraphics[height=0.49\textwidth, angle=270]{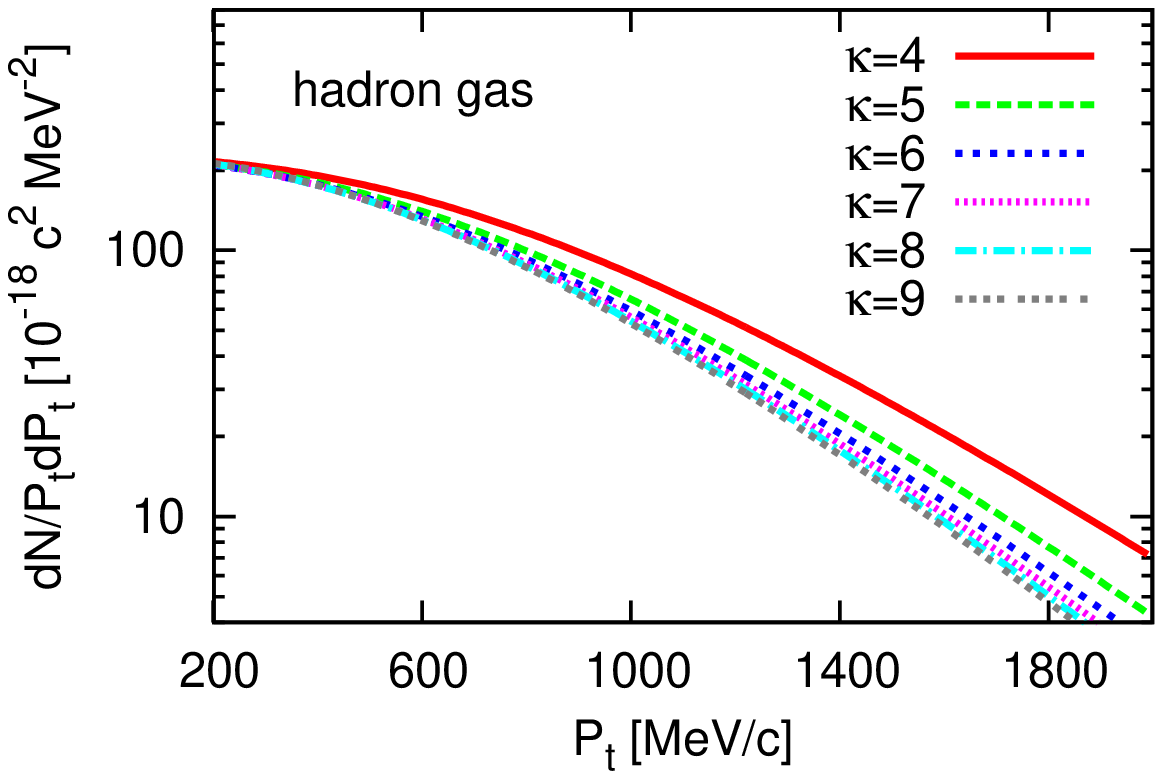}
 \includegraphics[height=0.49\textwidth, angle=270]{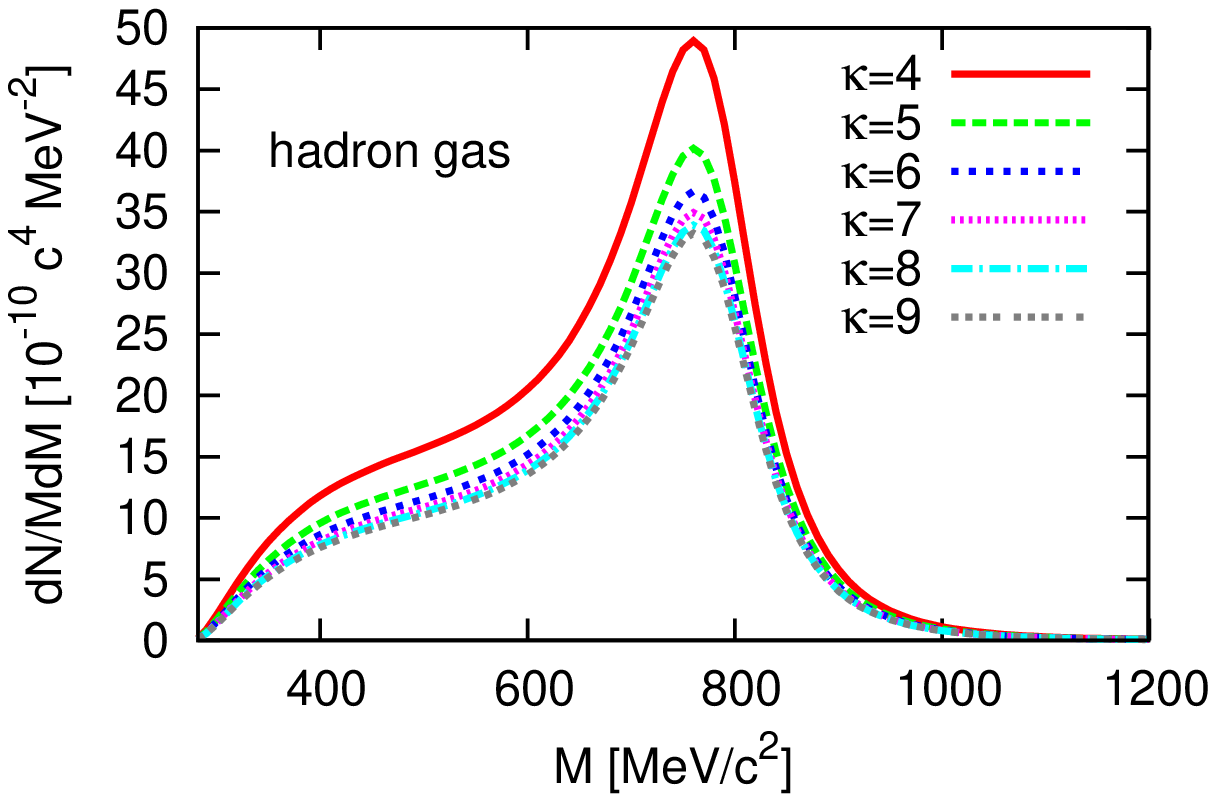}\\
 \end{center}
 \caption{Dilepton production for various equation of state parameters ($\kappa$), in case of quark
 annihilation (top) and pion annihilation in a hadron gas through the $\rho$ channel (bottom). Mass
 dependent curves are integrated out on $P_t$ (between 100 and 2000 MeV), while transverse momentum dependent results
 are taken at $M=1000$ MeV. These strongly depend on the Equation of State, the spectrum gets much steeper for
 increasing $\kappa$ values. The normalization changes as for a large $\kappa$, temperature changes slower
 as a function of proper time. Note also, that above $\kappa$ values of 5-6, dilepton invariant mass
 distributions are not very sensitive to the exact value of $\kappa$, see the lower right plot.}\label{f:kappadependence}
\end{figure}

EoS ($\kappa$) dependence of invariant mass ($M$) and the transverse momentum ($P_t$) distributions for 
both QGP and pion gas are shown in Fig.~\ref{f:kappadependence} (note that in the pion gas case, we take
into account the $\rho$ channel and its excited states ($\rho'$, $\rho''$), but for other contributions, one may
draw quite the same consequences). All distributions depend strongly on the 
Equation of State, in particular the $P_t$ spectrum gets much steeper for increasing $\kappa$ values. The 
absolute magnitude of the $M$ distribution changes with $\kappa$, as for a large $\kappa$, temperature
changes slower as a function of proper time. Hence (if the freeze-out temperature is fixed from hadronic data)
the system spends more time near the freeze-out temperature if $\kappa$ is large. In fact the experimental
data (of direct photons) supports large average $\kappa$ values ($\kappa=7.7$~\cite{Csanad:2011jq})
in QGP created at RHIC. As for the EoS in a pion gas, one may use lattice QCD values~\cite{Borsanyi:2010cj},
where a larger average is supported for subcritical temperatures. The used $\kappa$ values represent an average
Equation of State. Clearly, it changes throughout the evolution, as the temperature changes as well, but to have
an analytic description of dilepton creation, one has to assume a single average $\kappa$. Recently, solutions with a QCD
Equation of State (i.e. a $\kappa(T)$ function) were discovered~\cite{Csanad:2012hr}, but those
represent a constant temperature distribution, and are partly implicit, so we use the previously discussed solution here.
Also note, that above $\kappa$ values of 5-6, dilepton invariant mass distributions are not very sensitive to the
exact value of $\kappa$ in case of the hadron gas component, as shown on the lower right plot of Fig.~\ref{f:kappadependence}.

\begin{figure}
 \begin{center}
 \includegraphics[height=0.49\textwidth, angle=270]{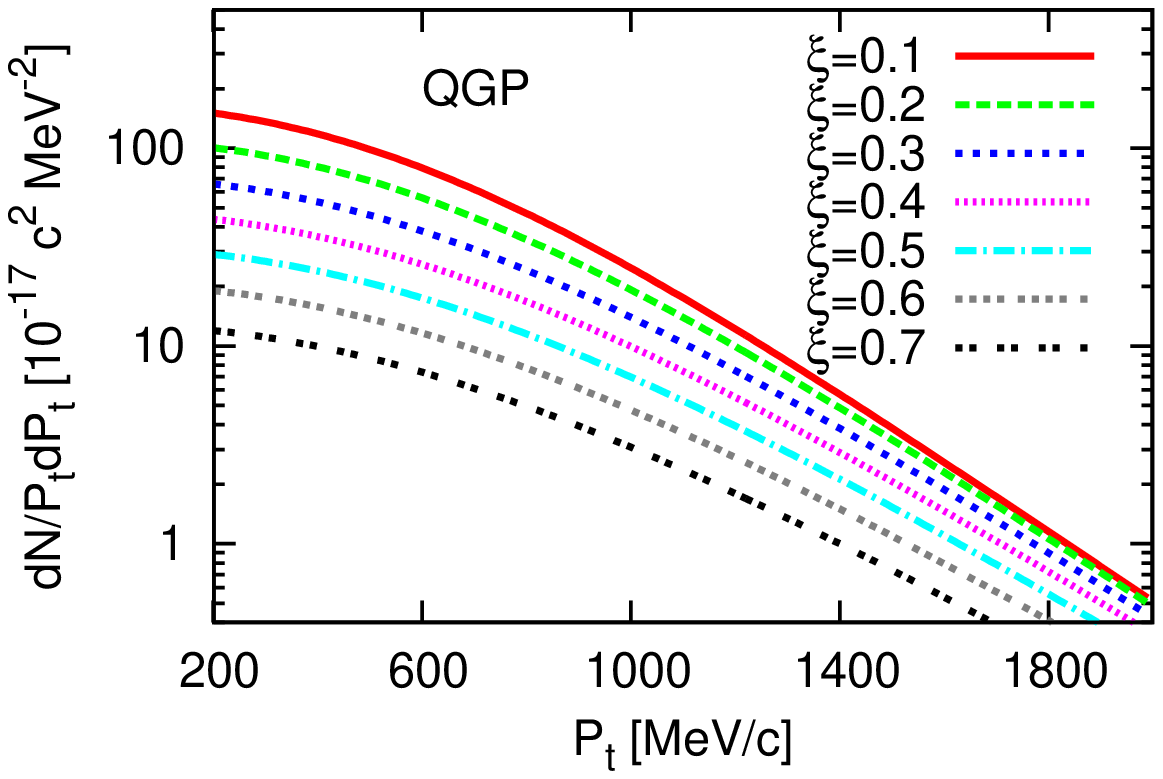}
 \includegraphics[height=0.49\textwidth, angle=270]{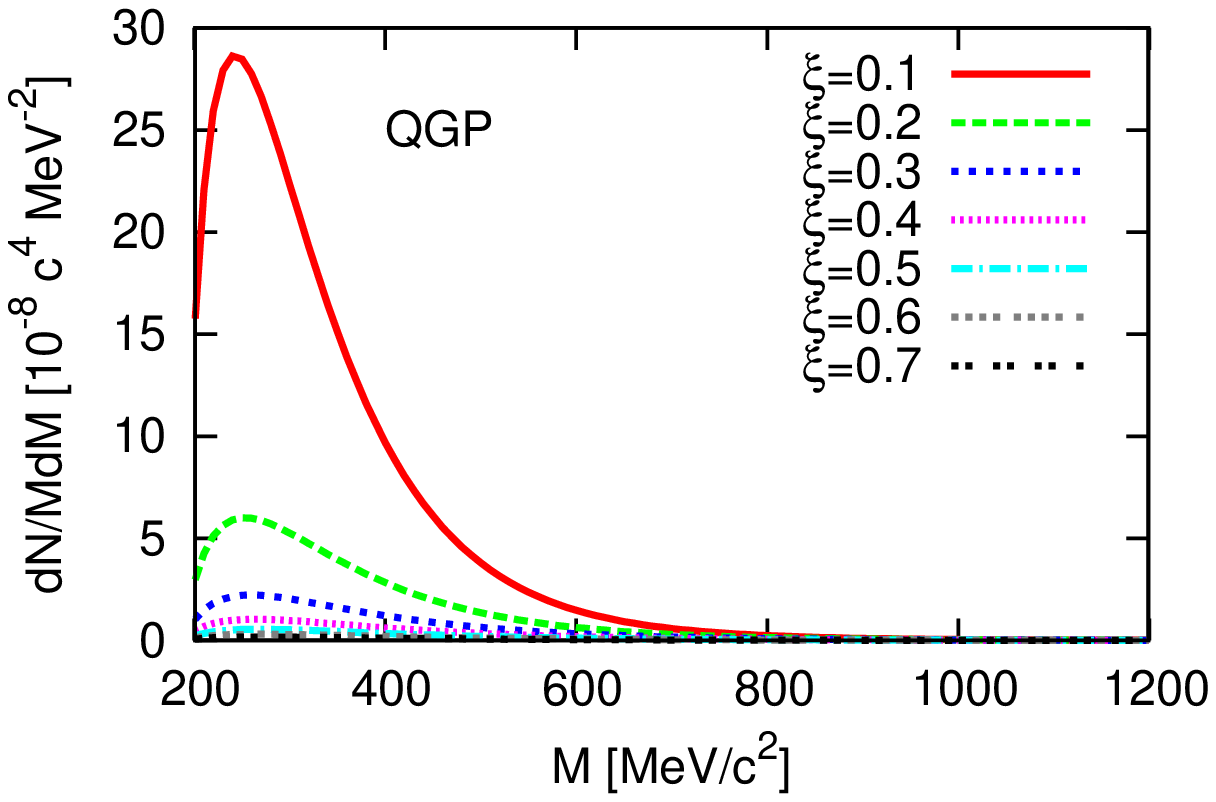}\\
 \includegraphics[height=0.49\textwidth, angle=270]{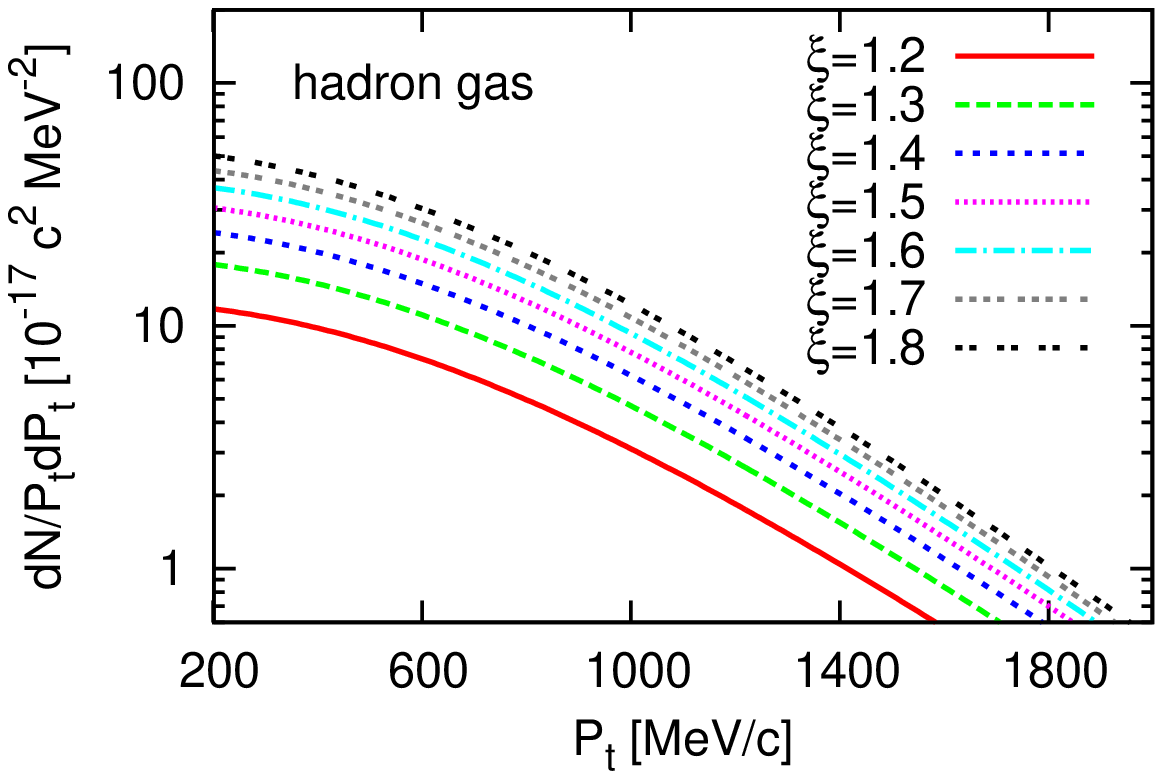}
 \includegraphics[height=0.49\textwidth, angle=270]{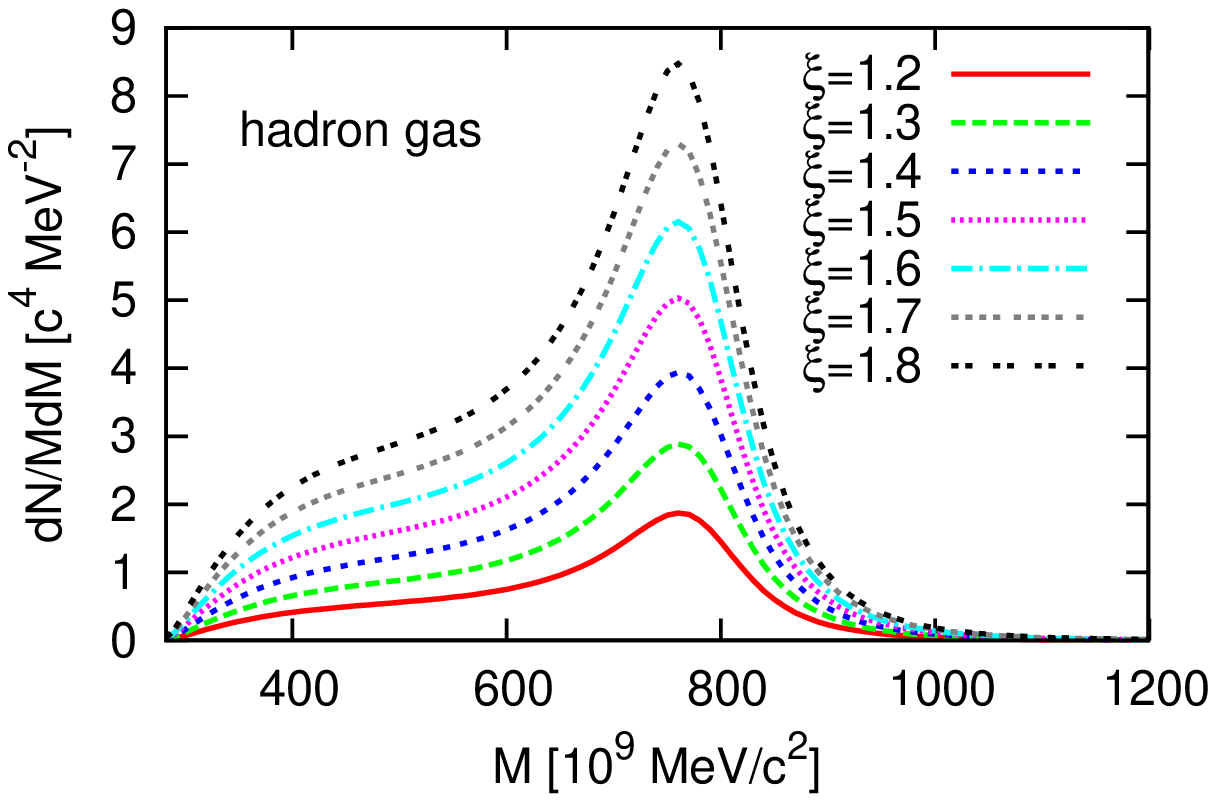}\\
 \end{center}
 \caption{Dilepton production dependence on the time evolution interval in case of quark annihilation (top) and
 pion annihilation (bottom), calculated similarly to Fig.~\ref{f:kappadependence}. In the first case, $\xi$ is defined
 as the initial dilepton production time divided by the quark-hadron transition (hadronic freeze-out) time. In the
 second case, $\xi$ is the ratio of the chemical freeze-out time and the hadronic freeze-out time.}\label{f:xidependence}
\end{figure}

We also calculated the dilepton production dependence on the evolution time interval, as clearly this determines
the absolute normalization of the results. Results on these are shown in Fig.~\ref{f:xidependence}. The curves
are labeled with changing $\xi$, which represents the ratio of the time integration limits (and the quark-hadron
transition time being always in the denominator, thus $\xi<1$ for QGP and $\xi>1$ for the pion gas). The results
show, that the longer the evolution lasts, the more dileptons are produced.

The above two paragraphs represent one of the key points of this analysis, since lepton production spans
the whole evolution of the Little Bangs created in heavy ion collisions. Thus, with a time dependent model,
one may extract information on the time evolution from dilepton observables. One of the most important
piece of information is the speed of sound, given through the Equation of State parameter $\kappa= 1/c_s^2$. 
In Ref.~\cite{Csanad:2011jq}, by comparing direct photons to hydrodynamic
calculations, we found a value of $c_s$=0.36$\pm$0.02, and an initial temperature (determined based on the
ratio of initial and final time, as well as on the freeze-out temperature) of $\sim$ 500 MeV. However,
dilepton production depends also on these numbers, and in Figs.~\ref{f:kappadependence},\ref{f:xidependence}
we analyzed this dependence. The Equation of State parameter $\kappa$ strongly influences the slope of the transverse
momentum distribution, but also the shape of the mass distributions.

\section{Comparision to data}
We compared our results to dilepton measurements on $\sqrt{s_{NN}}=200$ GeV RHIC Au+Au data of PHENIX~\cite{Adare:2009qk}. As mentioned
in the first paragraph of the previous section, we used the parameters from Ref.~\cite{Csanad:2009wc}
(where results from this model were compared to hadronic data), 
in particular the freeze-out time of 7.7 fm$/c$. From tuning our parameters to the data, we got for the central
initial QGP temperature (where thermal dilepton production starts) $T_{\rm ini}=270$ MeV, while in the hadron
gas, dileptons are produced until the central temperature drops to $T_{\rm final}=170$ MeV. These values
correspond to $\xi=0.49-1.0$ for the QGP phase and $\xi=1.0-1.6$ for the hadron phase. Note that the
length of the emissions correlates strongly with the weight of the given components. Also, these are values
for the central temperature, as our fireball is cooler in the outer regions (and temperature gradient is determined
by parameter $b$).
The EoS was determined in Ref.~\cite{Csanad:2011jq} (where results from this
model were compared to direct photon data), the value of $\kappa=7.7$ was used here as well. The results
in this range of 300 MeV $<M<$ 1800 MeV are not incompatible with the data, however, there is a
small excess around $M=500$ MeV, seen also if comparing to simulations based on p+p data~\cite{Adare:2009qk}.
Recently, it has been proposed that this excess may be attributed to the modification of the $\eta'$ mass~\cite{Vargyas:2012ci}.
Transverse momentum spectra are analyzed there with a modified $\eta'$ mass, and plugged in to the PHENIX dilepton
cocktail. Present results can be used to analyze possible mass and width modifications to describe the dilepton excess
at 500 MeV, based on modified momentum distributions; this is however outside the scope of present paper.
Note also, that the production mechanisms we included go to zero at low invariant masses, due to the vanishing 
phase-space of two-particle decays. However, three-particle decays of $\pi^0$, and $\eta$ or $\eta'$ account for
most of the dileptons produced at $M<200$ MeV. In order to get a simple, analytic and comprehensive result,
we sticked to the $\rho$, $\omega$ and $\phi$ contribution.

We also made a comparison to acceptance corrected thermal dilepton data from the SPS NA60
experiment~\cite{Damjanovic:2008ta} measured in 158 AGeV In+In collisions. Here we used similar parameters, however, a smaller
radial flow (0.64 instead of 0.84) and a smaller central temperature freeze-out ($T_{\rm fo}=140$ MeV instead
of 204 MeV)
were assumed (see e.g. Ref.~\cite{Ster:1999ib} for a motivation of these values). The dilepton production
starts also from a lower temperature, $T_{\rm ini}=200$ MeV here, and ends at
$T_{\rm final}=130$ MeV (corresponding to $\xi=0.4-1.0$ in QGP and $\xi=1.0-1.2$ in the hadron phase).
It is important to see, that when looking at the comparison, there is clearly a difference between the two
datasets: in case of PHENIX, quark annihilation plays the most important role, while for SPS, pion
annihilation through $\rho$ mesons is also important. Note though, that this contribution falls off faster
than the QGP contribution, as the latter has a higher temperature, so dileptons at a high invariant mass
have a higher probability of being created in QGP than in the hadronic phase. It is also clear, that even with the exchange
of higher mass mesons, data above $\approx$ 1 GeV$/c$ cannot be reproduced. This may be attributed to
other production mechanisms (not thermal production, as noted also in Ref.~\cite{Arnaldi:2006jq}), and the fact that we did not utilize in-medium
hadron spectral functions. If medium effects were considered in our calculation (which we did not do, as our goal was to obtain
a simple result and not go into the details of mass shifts and broadenings) then an enhancement would be expected around the
$\rho$ peak, and the data between $M_\rho$ and $M_\phi$ would be probably explainable. This would then
require a fine-tuning of the $\xi$ parameter for both QGP and the hadron phase, in order to describe the low
mass region as well. In such a detailed comparison, other processes and dilepton production channels
shall be taken into account, which we plan to accomplish in a subsequent analysis.

\begin{figure}
 \begin{center}
 \includegraphics[height=0.49\textwidth, angle=270]{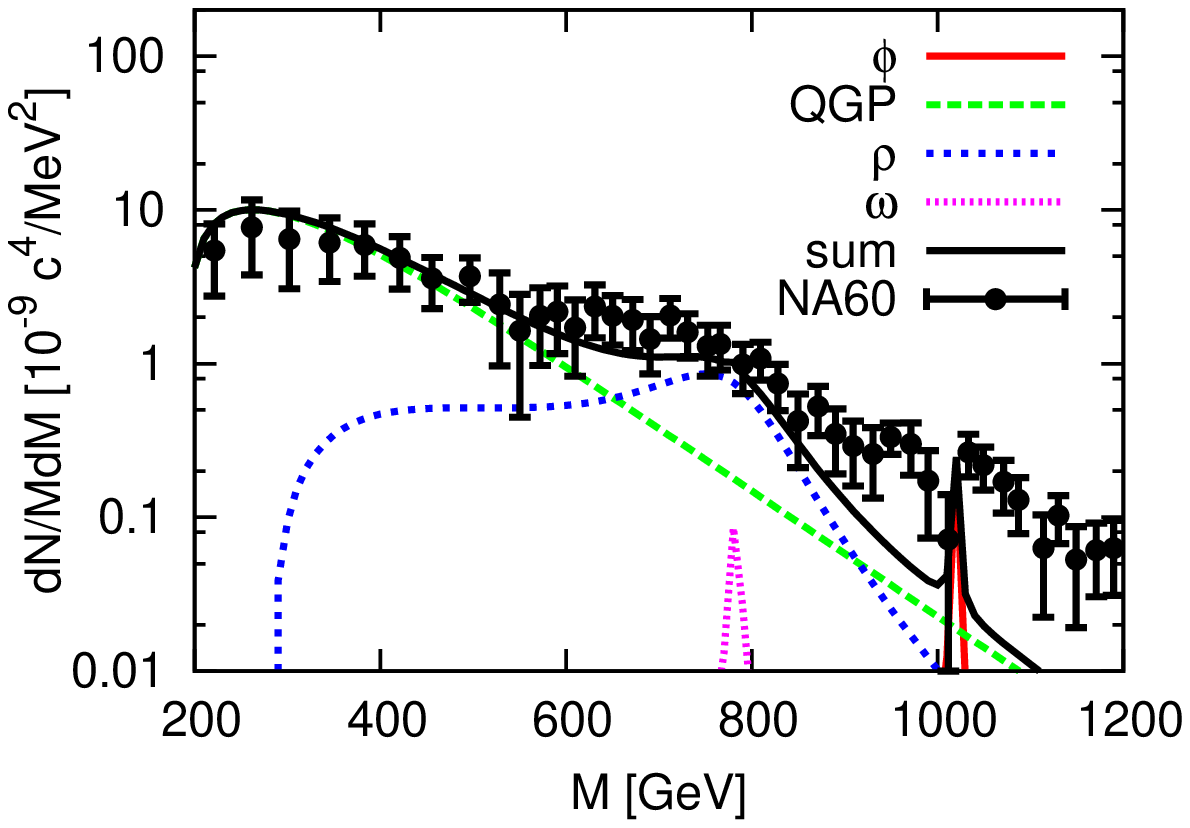}
 \includegraphics[height=0.49\textwidth, angle=270]{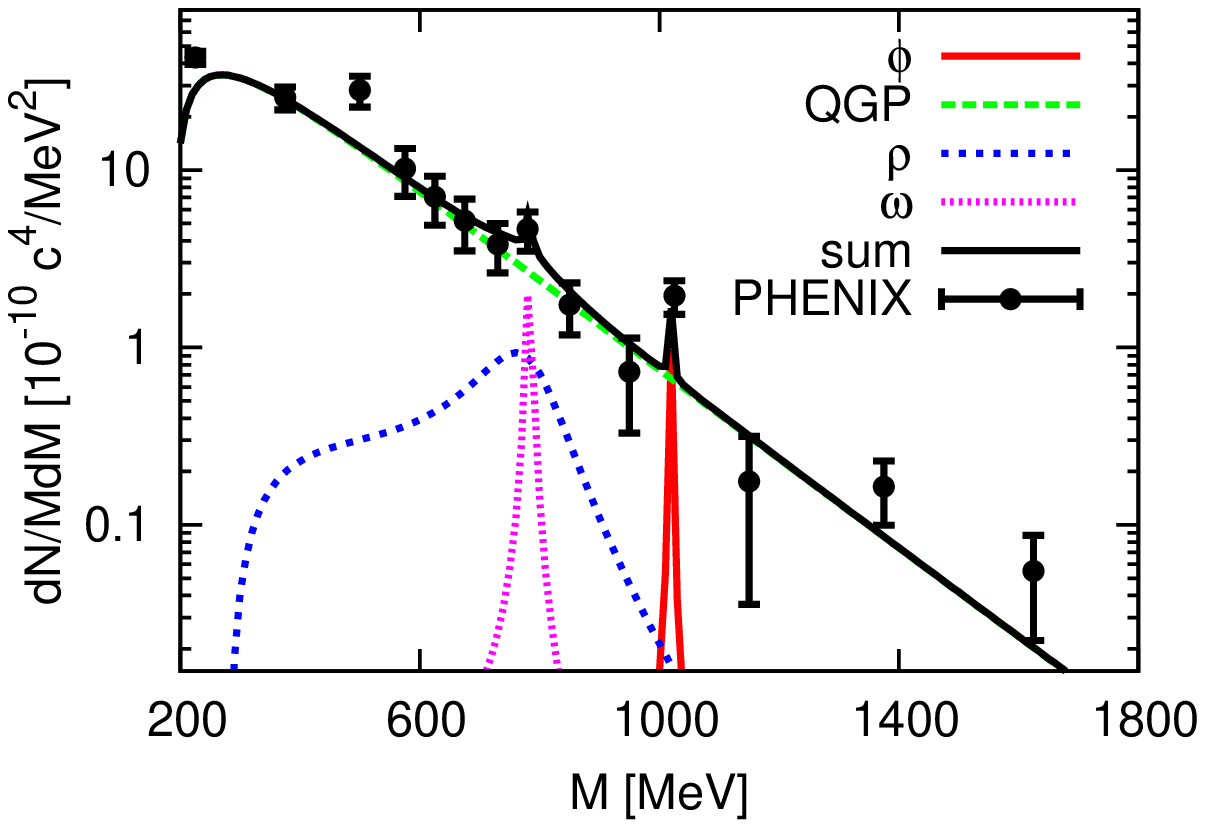}
 \end{center}
 \caption{Comparison with thermal dilepton production in $158$ AGeV In+In collisions of
 NA60~\cite{Damjanovic:2008ta} (0.4 to 0.6 GeV in $p_t$,
 acceptance corrected) and in $\sqrt{s_{NN}}=200$ GeV Au+Au collisions of  PHENIX~\cite{Adare:2009qk} (with PHENIX acceptance). At SPS,
 the $\rho$ contribution dominates dilepton production, but at some $M$ values there might be a QGP
 part of the data. However, at RHIC, predominantly QGP
 is the source of dileptons. Note that we have used vacuum parameters, not in medium values
 for the vector mesons, which may be the cause for the excess at 800-1000 MeV at SPS.
 At RHIC, the low mass dilepton excess is also not incompatible with this analysis, and may be due to
 the large contribution from low mass $\eta'$ production~\cite{Vargyas:2012ci}. Note however
 that in this simple analysis, we did not take into account the dilepton production via $\eta$
 and $\eta'$ mesons, neither the $\pi^{0}\rightarrow \gamma e^+ e^-$ channel, which are dominant
 for dilepton production at $M<200$ MeV.}\label{f:production}
\end{figure}

\section{Summary}
In high energy heavy ion collisions, particles are created via different production mechanisms. While hadrons are created at
the freeze-out of the medium, thermal photons and dileptons are constantly emitted from it. Hadrons thus reveal information
about the final state, whereas thermally radiated photons and dileptons carry information  about the whole time evolution. 
In this paper we calculated thermal dilepton production based on a hydrodynamical model.

The distinct feature of our paper is the analytic expression of the dilepton production based on 
an analytic 1+3d hydrodynamic solution, which depends only a few parameters, $\kappa$, $t_{\rm ini}$ and $t_{\rm final}$
(or the corresponding temperatures). The other parameters (such as transition time $t_0$ or transition temperature $T_0$,
transverse flow $u_t$) can be fixed from hadronic data. The dependence of the results on $\kappa$ (representing the EoS)
lifetime ratio $\xi$ has been analyzed. We found, that both the equation of state and the production time play an important 
role. Finally, we compared
our calculations to RHIC and SPS data, and found according to expectations, that in case of RHIC, quark annihilation plays
the most important role, while in case of SPS, a large portion of dileptons are produced in a hadron gas.
The RHIC dilepton data have been explained with the input parameters from earlier fits to RHIC direct photon
and hadron distributions.

\bibliographystyle{prlstyl}
\bibliography{../../../../Master}

\begin{thebibliography}{10}

\bibitem{Adcox:2004mh}
K. Adcox {\it et~al.}, Nucl. Phys. {\bf A757},  184  (2005)

\bibitem{Csorgo:2003ry}
T. Cs\"org\H{o} {\it et~al.}, Heavy Ion Phys. {\bf A21},  73  (2004)

\bibitem{Csanad:2009wc}
M. Csan\'ad and M. Vargyas, Eur. Phys. J. {\bf A44},  473  (2010)

\bibitem{Csanad:2011jq}
M. Csan\'ad and I. M\'ajer, Central Eur.J.Phys. {\bf 10},  850  (2012)

\bibitem{Csanad:2012hr}
M. Csan\'ad, M. Nagy, and S. L\"ok\"os, Eur.Phys.J. {\bf A48},  173  (2012)

\bibitem{Csanad:2009sk}
M. Csan\'ad, Acta Phys. Polon. {\bf B40},  1193  (2009)

\bibitem{Dusling:2006yv}
K. Dusling, D. Teaney, and I. Zahed, Phys.Rev. {\bf C75},  024908  (2007)

\bibitem{Ghosh:2010wt}
S. Ghosh, S. Sarkar, and J.-e. Alam, Eur.Phys.J. {\bf C71},  1760  (2011)

\bibitem{Kajantie:1986cu}
K. Kajantie {\it et~al.}, Phys. Rev. {\bf D34},  811  (1986).

\bibitem{Asakawa:1993kb}
M. Asakawa, C.~M. Ko, and P. L\'evai, Phys. Rev. Lett. {\bf 70},  398  (1993).

\bibitem{vanHees:2006ng}
H. van Hees and R. Rapp, Phys.Rev.Lett. {\bf 97},  102301  (2006)

\bibitem{Renk:2006qr}
T. Renk and J. Ruppert, Phys.Rev. {\bf C77},  024907  (2008)

\bibitem{Ruppert:2007cr}
J. Ruppert {\it et~al.}, Phys.Rev.Lett. {\bf 100},  162301  (2008)

\bibitem{Alam:2009da}
J.~K. Nayak {\it et~al.}, Phys.Rev. {\bf C85},  064906  (2012)

\bibitem{Song:2010fk}
T. Song, K.~C. Han, and C.~M. Ko, Phys.Rev. {\bf C83},  024904  (2011)

\bibitem{Kajantie:1986dh}
K. Kajantie {\it et~al.}, Phys. Rev. {\bf D34},  2746  (1986).

\bibitem{Beringer:1900zz}
J. Beringer {\it et~al.}, Phys.Rev. {\bf D86},  010001  (2012).

\bibitem{Borsanyi:2010cj}
S. Bors\'anyi {\it et~al.}, JHEP {\bf 11},  077  (2010)

\bibitem{Adare:2009qk}
A. Adare {\it et~al.}, Phys. Rev. {\bf C81},  034911  (2010)

\bibitem{Vargyas:2012ci}
M. Vargyas, T. Cs\"org\H{o}, and R. Vértesi, Central Eur.J.Phys. {\bf 11},  553
   (2013)

\bibitem{Damjanovic:2008ta}
S. Damjanovic, J.Phys. {\bf G35},  104036  (2008)

\bibitem{Ster:1999ib}
A. Ster, T. Cs\"org\H{o}, and B. L\"orstad, Nucl. Phys. {\bf A661},  419
  (1999)

\bibitem{Arnaldi:2006jq}
R. Arnaldi {\it et~al.}, Phys.Rev.Lett. {\bf 96},  162302  (2006)

\end{thebibliography}

\end{document}